\documentclass[showkeys,aps]{revtex4}

\usepackage{graphicx}
\usepackage{dcolumn}
\usepackage{bm}
\usepackage{amsmath}
\usepackage{amssymb}
\usepackage{latexsym}
\usepackage{epsfig}
\usepackage{amsbsy}
\usepackage{array}
\usepackage{amssymb}
\usepackage{setspace}
\usepackage{bm}
\usepackage{textcomp}
\usepackage{subfig}
\usepackage{epstopdf}
\usepackage{float}
\usepackage{color}

\begin{document}

\title{Phase transitions of antiferromagnetic Ising spins on the zigzag surface of an asymmetrical Husimi lattice}

\author{Ran Huang \footnote{Correspondence to: ranhuang@sjtu.edu.cn}}

\affiliation{State Key Laboratory of Microbial Metabolism and School of Life Sciences and Biotechnology, Shanghai Jiao Tong University, Shanghai 200240, China}
\affiliation{Department of Materials Technology and Engineering, Research Institute of Zhejiang University-Taizhou, Taizhou, Zhejiang 318000, China}

\author{Purushottam D. Gujrati \footnote{Correspondence to: pdg@uakron.edu}}
\affiliation{The Department of Physics; The Department of Polymer Science, The University of Akron, Akron, OH 44325}

\date{\today}

\begin{abstract}

 An asymmetrical 2D Ising model with a zigzag surface, created by diagonally cutting a regular square lattice, has been developed to investigate the thermodynamics and phase transitions on surface by the methodology of recursive lattice, which we have previously applied to study polymers near a surface. The model retains the advantages of simple formulation and exact calculation of the conventional Bethe-like lattices. An antiferromagnetic Ising model is solved on the surface of this lattice to evaluate thermal properties such as free energy, energy density and entropy, from which we have successfully identified a first order order-disorder transition other than the spontaneous magnetization, and a secondary transition on the supercooled state indicated by the Kauzmann paradox.

\end{abstract}
\keywords{Zigzag Surface, Asymmetrical Husimi Lattice, Phase Transitions on Surface}

\maketitle

\section{Introduction}

The recursive lattice has been a classical methodology in statistical physics for several decades since its invention by Bethe \cite{Bethe} and Husimi \cite{Husimi}. With its impressive advantages of exact calculation and simple formulation due to the feature of recursive structure, many statistical or physical problems can seek an alternative but reliable solution from this modeling method, e.g. the travelling-salesman problem \cite{Krauth_Mezard}, the K-satisfiability \cite{K_satisfiability}, the glass transition \cite{Mezard_Parisi} and so on. For most cases, especially when involving complex structures, it is impossible to solve arbitrary models on a square lattice so one is forced to find approximated solutions. Usually, one attempts to solve the model in a mean-field approximation, while Gujrati \cite{pdg_prl} established that recursive lattice solutions are more reliable than the mean-field solutions, especially for the antiferromagnetic models. 

On the other hand, the recursive feature of such lattices implies that the system is naturally homogeneous and suitable to describe bulky systems. Our group has already used the Bethe lattice to study inhomogeneous structures near a surface \cite{Mukesh1,Mukesh2,Mukesh3}. Here we extend to a 1D/2D hybrid recursive lattice to study the phase transitions of Ising model on surface/interface, which is nonetheless still based on a symmetrical structure \cite{ctp_surface}.  

In this work, we have constructed an asymmetrical 2D recursive lattice with a 1D surface. By fractaclizing the analog of a diagonally cut regular square lattice, we can obtain a simple model with partial Husimi trees hung on a zigzag surface. Since this zigzag structure is infinite and homogeneous, it can be handled by recursive calculation \cite{zigzag,triangle}, with approximating partial Husimi trees to be constant statistical contributions, which was derived from previous classical works, the exact calculation of this model appears to be feasible as like other Bethe-like models. An antiferromagnetic Ising model have been solved on the lattice, and we can locate the 1st-order phase transition and the Kauzmann paradox in the $ \pm1 $ spin system. 
     
\section{Lattice and model}

\begin{figure}
 	\centering{
 		\includegraphics[width=0.6\textwidth]{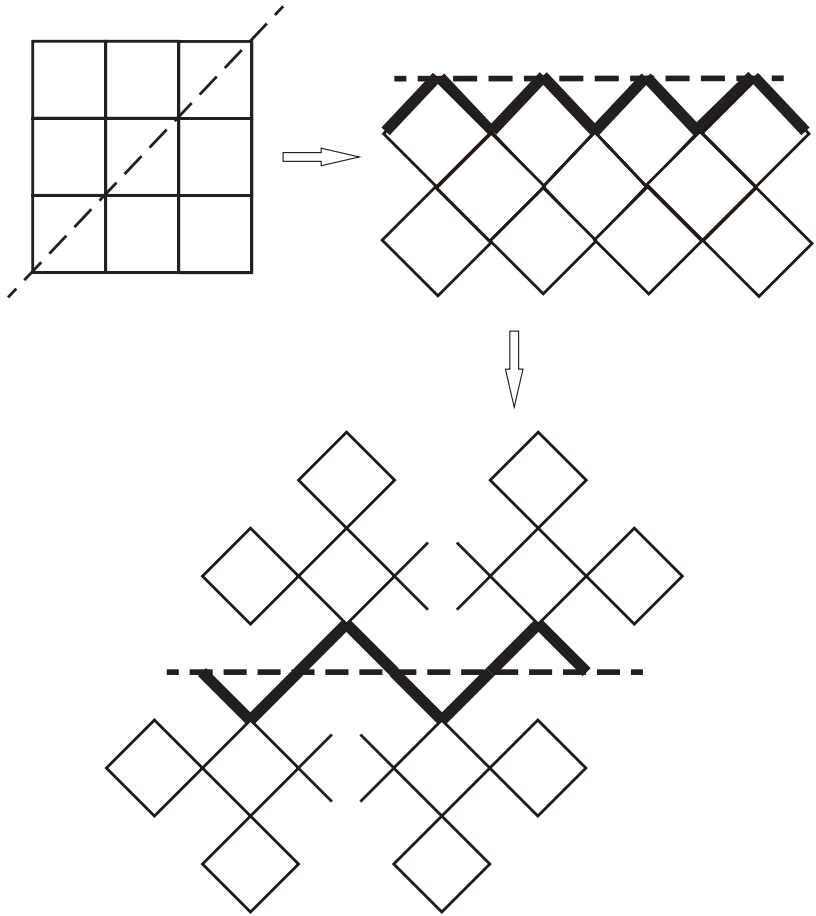}
 		\caption{The diagonal cutting on a regular square lattice to obtain a zigzag surface, and the construction of recursive lattice on the zigzag surface. The thick line presents the surface edge.}
 		\label{fig1}
 	}
 \end{figure}

Conventionally, the surface of a 2D square lattice is just the 1D line confining the bulk with a lower coordination number of 3. Imagine we cleave the lattice along square diagonals, a zigzag surface can be generated for the one half bulk, and the basic unit of this lattice is square in the bulk and triangle on the surface. The triangle unit is surrounded by two identical triangles and a bulk square, thus a straight connection of triangle units can be adopted to represent the surface. This triangle chain is infinite and has the recursive feature required for calculation. Considering that bulk parts are merely structures hung on the zigzag line, on each triangle we may attach an infinite partial Husimi tree to represent the bulk part. The scheme of fractalizing the halved structure to be a recursive lattice is shown in Fig. \ref{fig1}. 
 
\begin{figure}
	\centering{
		\includegraphics[width=0.6\textwidth]{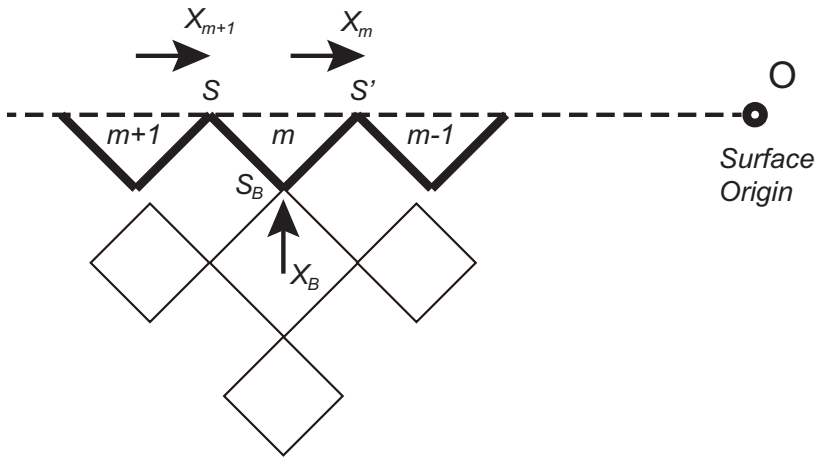}
		\caption{The sites labeling and calculation scheme on ZSRL. Starting from a random point on the surface, the recursive approach is proceed to an imaginary origin.}
		\label{fig2}
	}
\end{figure}

 Since this zigzag surface recursive lattice (ZSRL) has coordination of 4 inside the bulk and alternating 2 or 4 (averagely 3) on the surface, we hope that it is a good approximation to the regular square lattice with a diagonal boundary. The sites labeling is shown in Fig. \ref{fig2}: the vertices on surface triangle are labeled as $ S $ and $ S' $, where $ S' $ marks the site closer to an imaginary origin point on the surface, following the direction of recursive calculation. Note the $ S' $ in one triangle is $ S $ in the unit of lower level while the origin is marked as level 0. The base site where links the surface triangle and bulk tree is labeled as $ S_{B} $.

Corresponding to the notation of neighbor interaction $J$, diagonal interaction $ J_{P} $, and three spins interaction (triplet) $ J' $, on the surface unit we label a bar on each term to differ them from bulk parameters. Then there are two neighbor interactions $ \bar{J} $, one diagonal interaction $ \bar{J}_{P} $, and one triplet ($ \bar{J'} $) in the Hamiltonian of one triangle unit $\alpha$ is

\begin{equation}
e_{\alpha}=-\bar{J}(S_{B}\cdot S+S_{B}\cdot S')-\bar{J}_{P}\cdot S\cdot S'-\bar{J'}\cdot S_{B}\cdot S\cdot S'-H(S+S_{B}), \label{e_alpha}%
\end{equation}
where $ H $ is the magnetic field applied to each spin. Note the $H$ of site $ S' $ is not included in the last term, because it will be counted in the unit of next level. In eqn.\ref{e_alpha}, a negative value of $\bar{J}$ will raise the lowest energy with different neighboring spins states, the system is thereafter the antiferromagnetic case, or vice versa. The detailed effects of energy parameters will be discussed in the section \ref{effect}.

\section{Calculation}

\subsection{Solutions on the Surface}

The general calculations of Ising model on Husimi square lattice can be found in previous works \cite{pdg_prl,ctp1}. Here we follow the similar method of partial partition function (PPF) to achieve the solution $ \bar{x} $, which presents the probability of one site to be occupied by + spin on the surface site. The triangle unit has $ 2^3 = 8 $ possible configurations, four of which are with the base site $ S' = +1 $ and the others are with $ S' = -1 $. We can derive two PPFs of the triangle on a level $ m $ from PPFs of the higher level $m+1$, the contribution from bulk $ Z_{B}(S_{B}) $, and the local weight $ w(\Gamma) $:

\begin{equation}
Z_{m}(+)  =\underset{\Gamma=1}{\overset{4}{\sum}}Z_{m+1}(S_{m+1})Z_{B}(S_{B})w(\Gamma), \label{PPF+_Recursion}%
\end{equation}

\begin{equation}
Z_{m}(-)  =\underset{\Gamma=5}{\overset{4}{\sum}}Z_{m+1}(S_{m+1})Z_{B}(S_{B})w(\Gamma),
\label{PPF-_Recursion}%
\end{equation}
where $\Gamma$ is the index of configuration.  Note that the total partition function PF of the entire system at the origin site $ S_0 $ is given by
\begin{equation}
Z_{0}  =Z_{0}(+)^2e^{-\beta H} + Z_{0}(-)^2e^{\beta H}, \label{PF}%
\end{equation}
this relationship is fundamental when we derive thermal quantities, e.g. free energy from the PPF.

We introduce ratios 
\begin{equation}
\bar{x}_{m}=\frac{Z_{m}(+)}{Z_{m}(+)+Z_{m}(-)},\text{ }\bar{y}_{m}=\frac{Z_{m}(-)}%
{Z_{m}(+)+Z_{m}(-)} \label{Ratios}%
\end{equation}
as the solutions on surface at level $m$. By denoting 
\begin{equation}
\bar{z}_{m}(S_{m})=\left\{
\begin{array}
[c]{c}%
\bar{x}_{m}\text{ if }S_{m}=+1\\
\bar{y}_{m}\text{ if }S_{m}=-1
\end{array}
\right.  , \label{General Ratio}%
\end{equation}
we have $Z_{m}(+)=B_{m}\bar{x}_{m}$ and $Z_{m}(-)=B_{m}\bar{y}_{m}$ with $ B_{m}= Z_{m}(+) +Z_{m}(-) $ , then with Eq.\ref{PPF+_Recursion} and \ref{PPF-_Recursion} we have 

\begin{equation}
\bar{z}_{m}(S_{m})=\frac{B_{m+1}B_{B}}{B_{m}}\underset{\Gamma}{\sum}%
\bar{z}_{m+1}(S_{m+1})z_{B}(S_{B})w(\Gamma).
\end{equation}
By introducing a set of polynomials

\begin{equation}
Q_{m+}(\bar{x}_{m+1},x_{B})=\underset{\Gamma=1}{\overset{4}{\sum}}%
\bar{z}_{m+1}(S_{m+1})z_{B}(S_{B})w(\Gamma),
\end{equation}

\begin{equation}
Q_{m-}(\bar{y}_{m+1},y_{B})=\underset{\Gamma=5}{\overset{4}{\sum}}%
\bar{z}_{m+1}(S_{m+1})z_{B}(S_{B})w(\Gamma),
\end{equation}

\begin{equation}
Q_{m}(\bar{x}_{m+1},x_{B})=Q_{m+}(\bar{x}_{m+1},x_{B})+Q_{m-}(\bar{y}_{m+1},y_{B})=
\frac{B_{m}}{B_{m+1}B_{B}},
\end{equation}
we can derive the solution $ \bar{x}_{m} $ at the $ m $th level as a function of the ratio $ \bar{x}_{m+1} $ on higher level and the bulk solution $ x_{B} $:

\begin{equation}
\bar{x}_{m}=\frac{Q_{m+}(\bar{x}_{m+1},x_{B})}{Q_{m}(\bar{x}_{m+1},x_{B})}.
\label{solution}%
\end{equation}

Taking $ x_{B} $ as the solution on a joint site between surface triangle and bulk square, imagine we start the recursive calculation from a position deep in the bulk then approach the thermodynamic contributions to the surface, thereby $ x_{B} $ is simply the fix-point solution of Husimi lattice. In this way we can count the contribution of the infinite bulk tree
as a constant input and focus on the recursive approach of $ \bar{x} $ along the surface. It should be addressed that this setup of constant $ x_{B} $ ignores the possible backward effect from surface to the bulk, which may bias the numerical value of $ x_{B} $. However this approximation is acceptable to make the model simple and solvable. 

With a constant $ x_{B} $, by Eq.\ref{solution} we can recursively calculate the solution on surface for a number of iterations until we reach a fix-point solution. The form of Eq.\ref{solution} implies that, regardless of the system being antiferro- or ferromagnetic, a uniform 1-cycle solution is expected on the surface, while antiferromagnetic Ising model presents an alternating 2-cycle solution as the ordered state \cite{ctp1}. A negative neighbor interaction $ \bar{J} $  prefers to anti-align the $ S $ vs $ S_{B} $ and $ S' $ vs $ S_{B} $ pair, unless we set the diagonal interaction $ \bar{J}_P $ also to be negative and large enough to outweigh $ \bar{J} $, the system will prefer the same spin states on $ S $ and $ S'$. 

Recall that bulk solutions can be either a 1-cycle solution to present the metastable state, or a set of 2-cycle solutions as the ordered state \cite{ctp1}. Taking the 1-cycle $ x_{B} $, which is usually 0.5 with $H=0$, we will obtain a fixed $ \bar{x}=0.5 $ also corresponding to a metastable surface; for the 2-cycle solutions, we can substitute either fixed solution as  $ x_{B} $ and it will affect the calculation on $ \bar{x} $ to bias the surface fix-point solution to 0 or 1. For example, due to antiferromagnetism, at $T=0$ with $ H = 0 $, we will have 0 and 1 solutions in the bulk, then if we take $ x_{B}=0 $, the surface solution will be captured as $ \bar{x}=1 $, and vice versa. Our results confirmed that the thermodynamics calculated based on either selection are identical. 

\begin{figure}
    \centering{
    	\includegraphics[width=0.6\textwidth]{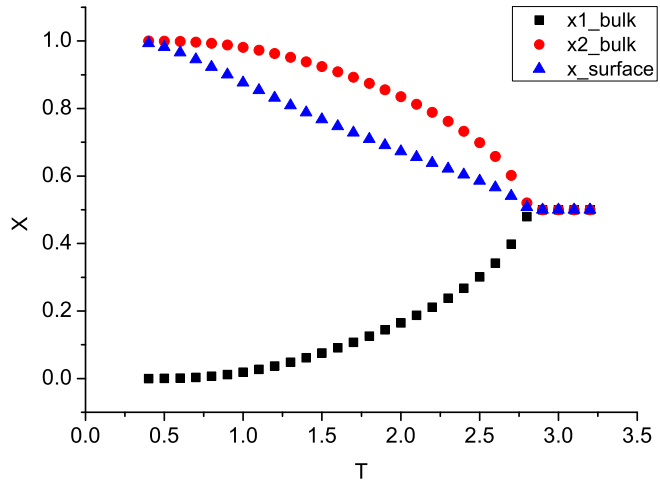}
    	\caption{Solution on the surface and its comparison to the 2-cycle bulk solutions.}
    	\label{fig3}
        }
\end{figure}

Fig. \ref{fig3} shows the ordered solution on the surface with comparison of 2-cycle bulk solutions, the energy terms are set as $ J=-1 $, $ \bar{J}=-1 $, and others to be 0. The metastable $ 0.5 $ solution is not presented in Fig. \ref{fig3}. For convenience, we are still going to call the stable solution and corresponding thermodynamics ``2-cycle" in the following discussion, although both stable and metastable $\bar{x}$ are actually in 1-cycle form for ZSRL. 

It can be observed that at high $ T $ all the solutions are 0.5, that spins anywhere have an equal probability to be $\pm1$. Below the Curie point, the spins undergoes self-magnetization (even without an external field $ H $), and the neighboring spins prefer the $+/-$ alternating arrangement, which is the lowest energy state. The value of 2-cycle $x_B$ of bulk Ising model is referred from previous reports \cite{ctp_surface,ctp1}. Along with the bulk solutions, the spins on surface also present preference on unitary direction under the bulk Curie point, however we will show that the actual phase transition occurs far below the $ T_C $(bulk).

One more thing should be addressed here. We know from exact solutions that for the square lattice the self-magnetization occurs at $ T_C = 2/log(1+\sqrt{2}) \sim 2.27 $. In the recursive lattice, the bulk transition temperature is 2.8 as shown in Fig. \ref{fig3}, which is not all that close to the known $ T_C $, but closer than the results of mean-field theory. We believe this overestimate is acceptable. The aspect that the RL method providing results between the `real' exact solution and the mean-field theory seems to be generic. Similar results had also been observed in other works, e.g. the 3D cube lattice \cite{ctp1}. However, the nature of this overestimate has not been investigated yet.

\subsection{Free Energy Calculation}

\begin{figure}
	\centering{
		\includegraphics[width=0.4\textwidth]{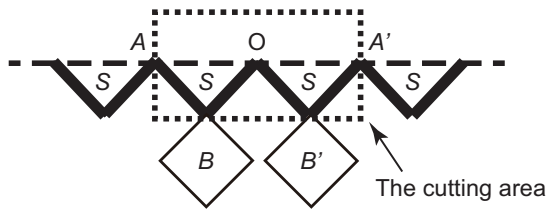}
		\caption{The cutting scheme around the surface origin $ O $ for free energy calculation.}
		\label{fig4}
	}
\end{figure}

We follow the Gujrati trick to to calculate the free energy of a local area by recursive approach \cite{pdg_prl,ctp1}. The scheme will be briefed here as shown in Fig. \ref{fig4}. Due to the uniform structure and solution on the surface, we can randomly select a site as the origin point $ O $. The local area is chosen to be two triangle units joint on the origin. Imagine we cut off two sub-trees contributing to the point $ A $ and $ A' $ then rejoin them together to make an identical but smaller ZSRL, and hook up two partial bulk trees hung on $ B $ and $ B' $ together to make a full Husimi square lattice, therefore we have 
$
F_{total} = F_{local} + F_{bulk} + F_{smaller}
$. With the definition of Helmholtz free energy $F=-k_bTlogZ$, the $F$ per site in the local area is given as:
\begin{equation}
F_{site}=-\frac{1}{3}Tlog(\dfrac{Z_0}{Z_1\cdot Z_B}),
\end{equation}
since there are three full sites in the local area (on whole site $ S_0 $ and four half-shared sites $S_1$, $S_B$), and $k_B$ is normalized to be 1.
 
Recall Eq.\ref{PPF+_Recursion}, \ref{PPF-_Recursion} and \ref{PF}, it is easy to break down the local free energy into a function of PPFs, then from the relations between $Z(\pm)$, $Q$ and $x$ derived in the previous section, we can calculate the above function as

\begin{equation}
F_{site}=-\frac{1}{3}Tlog(\dfrac{{Q_0}^2}{{x_B}^2e^{\beta H}+(1-x_B)^2 e^{-\beta H}}),
\end{equation}
then the entropy and energy (per site) can be easily achieved by $ S=-dF/dT $ and $E=F+TS$.

\section{Results and discussion}

\subsection{The thermodynamics and transitions on the surface}

\begin{figure}
    \centering{
    	\includegraphics[width=0.6\textwidth]{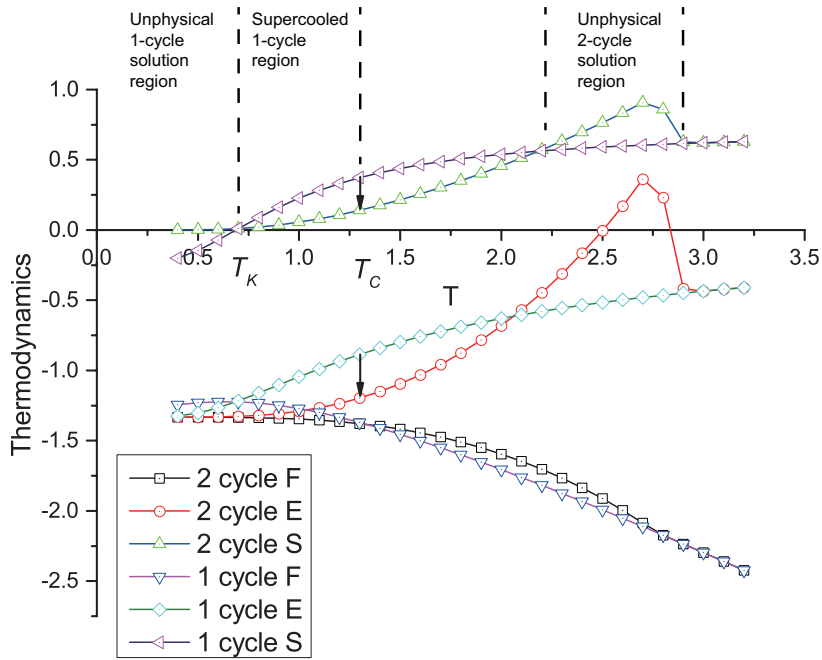}
    	\caption{The thermodynamics behaviors on the surface with $ J = -1 $, $ \bar{J} = -1 $ and other parameters to be 0.}
    	\label{fig5}
        }
\end{figure}

The thermal behaviors of the reference setup with $ J = -1 $, $ \bar{J} = -1 $ and other parameters to be 0 is shown in Fig. \ref{fig5}. The free energy of two solutions differ at $ T = 2.8 $. Usually this bifurcation indicates the spontaneous magnetization (Curie point) in normal spin models. However, here on the surface the magnetization does not bend the free energy of alternating spins arrangement below the disordered 0.5 solution but upward, which is unphysical with a sharply increased entropy. Therefore the 2-cycle solution at this point is only a numerical existence and the system must follow the curve of 1-cycle solution, until a cross point is reached at $ T = 1.33 $,  where the system makes a transition from the amorphous to the crystalline ordering, i.e. the order-disorder transition at critical temperature $T_C$. Unlike the conventional self magnetization where entropy is continuous, the entropy here must undergo a discontinuous jump as like a first order transition. In this way, we may conclude that the critical order-disorder transition on the surface is not the Curie point, but much lower than that of the bulk. This can be understood as that, due to the asymmetry and smaller coordination numbers on the surface, spins on the outer layer are less dragged by the bulk portions. Below the real $ T_C $(bulk) and the apparent Curie point on the surface, the spins on surface can maintain their ``melted" state in a deep temperature region, and this does not fall into the concept of supercooled liquid, since the corresponded crystal state in particular temperature region is unfeasible. Intuitively this phenomenon may easily refer to the analog of ice premelting \cite{premelting}; nonetheless this work only focuses on the preliminary modeling and we are not going to further expand this point.

For the 1-cycle solution below $ T_C $, the system can undergo cooling without any phase transition and shares features of a supercooled state. With further cooling process, the entropy of 2-cycle solution approaches to zero, while the entropy of 1-cycle becomes negative at $ T = 0.69 $, which is the Kauzmann paradox at $T_K$ \cite{ctp1}. 

\begin{figure}
	\centering{
		\includegraphics[width=0.6\textwidth]{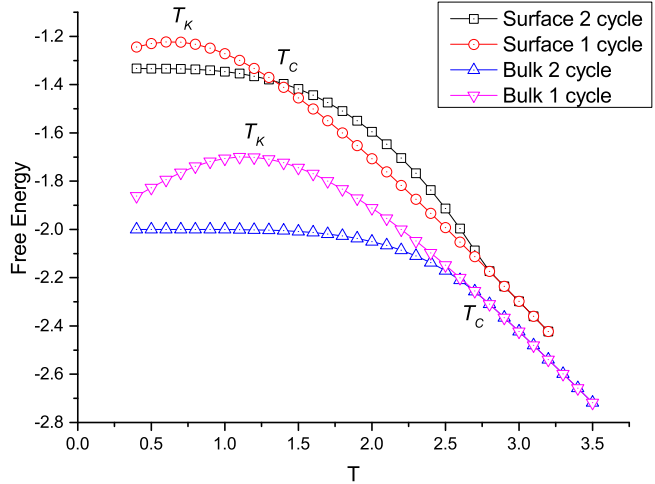}
		\caption{The free energy comparison of Husimi bulk system and ZSRL $ J = -1 $, $ \bar{J} = -1 $ and other parameters to be 0.}
		\label{fig6}
	}
\end{figure}

The result indicates that a free surface dramatically decreases the transition temperatures. Fig. \ref{fig6} shows the free energy comparison of Husimi bulk system and ZSRL. This observation agrees with others' work on phase transitions on the surface or thin film, for example the glass transition of polymer system in confined geometry \cite{Mukesh1,Mukesh2,Mukesh3,Forrest}. The fact that similar reduction can be observed in our monoatomic model implies that the lower transition temperature on surface/thin film basically originates from the dimension reduction and less interaction constraints. 

\subsection{The effects of secondary energy parameters} \label{effect}

Besides mathematical curiosity, there is always a primary expectation to describe and study real systems for establishing a theoretical model. In this way, other than the interaction $J$ and $\bar{J}$ between the nearest neighbors, further interactions are included to make the model more versatile to describe various systems, as shown in Eq. \ref{e_alpha}. With adjustable combinations of energy parameters setup, we can manipulate the thermal behavior of system and the transition temperatures, to better match the reality in particular situations. In another word, more adjustable parameters may serve as useful tools for the theoretical modeling to be correlated with experimental parameters. By the control of $J=-1$ and other parameters to be 0 inside the bulk, we explored the effects of $ \bar{J} $,  $ \bar{J}_P $ and $ \bar{J'} $. The secondary parameters are expected to be either comply or compete with $J$, and some interesting phase behaviors are found with particular setups. 

\subsubsection{The surface nearest-neighbor interaction  $ \bar{J} $}

Considering the feature of asymmetry on the boundary, we set the nearest-neighbor interaction on the surface, denoted as $\bar{J}$, to be differentiated and adjustable from the $J$, to make the model capable to describe some particular situations, e.g. the surface tension. The transition temperatures with $ J=-1 $, other parameters to be 0, and  $ \bar{J}  = -0.5$, $ -0.7 $, $ -0.9 $, $ -1.1 $, $ -1.3 $, and $ -1.5 $ are shown in table \ref{tab1}.

As negative $\bar{J}$ complies with antiferromagnetic setup $ J=-1 $, the larger absolute value of $\bar{J}$ makes the system more stable and increases both critical and ideal glass transition temperature; and the relative length of supercooled region, indicated by the ratio $T_{C}/T_{K}$, gradually increases. However for the $\bar{J} =-0.5$ decreased from $-0.7$, both $T_{C}$ and $T_{K}$ falls while the latter changes more dramatically, raising a much larger supercooled region. This critical phenomenon observes that a loosen surface with too weak interactions is easier to be supercooled.

\begin{table}[]
\caption{The transition temperature variations with different  $ \bar{J} $.}
\begin{center}
\begin{tabular}{llll} 
\hline\hline $ \bar{J} $ & $T_{C}$& $T_{K}$ & $T_{C}/T_{K}$\\
\hline -0.5 &  0.85 & 0.40 & 2.13 \\
     -0.7  & 1.10 & 0.60 & 1.83\\
     -0.9 & 1.20 & 0.63 & 1.90 \\
     -1   & 1.33 & 0.69  & 1.93\\
     -1.1 & 1.50 & 0.77 & 1.95 \\
     -1.3 & 1.80 & 0.90 & 2.00 \\    
     -1.5 & 2.10 & 1.01 & 2.08 \\   
\hline \hline
\end{tabular}
\end{center} \label{tab1}
\end{table}

\subsubsection{The diagonal interaction $ \bar{J}_P $}

The diagonal interaction  $ \bar{J}_P $ between the two top sites in the triangle unit is the only competition to
the nearest-neighbor $ \bar{J} $. Transition temperatures with $ \bar{J}_P   =   \pm0.2 $, $ \pm0.4 $ and $ 0.6 $ are summarized in table \ref{tab2}. As expected, same polarity of $ \bar{J}_P $ and $ \bar{J} $ obstacles the ordering degree with reduced $T_{C}$ and $T_{K}$, while positive $ \bar{J}_P $ complies with $ \bar{J} $ and gives higher $T$s and larger supercooled region. A similar phenomenon of sharply reduced $T$s and enlarged $T_{C}/T_{K}$ ratio is also observed with $ \bar{J}_P=-0.4 $, which implies that a vigorous competition between $ \bar{J}_P $ and $ \bar{J} $ has the same effect of weak $ \bar{J} $ to make the surface preferring supercooled state.

\begin{table}[]
\caption{The transition temperature variations with different $ J_{P} $.}
\begin{center}
\begin{tabular}{llll} 
\hline\hline $ \bar{J}_P $ & $T_{Cm}$& $T_{K}$ & $T_{C}/T_{K}$\\
\hline -0.4 &  0.85 & 0.40 & 2.13 \\
     -0.2  & 1.10 & 0.60 & 1.83\\
     0   & 1.33 & 0.69  & 1.93\\
     0.2  & 1.54 & 0.75  & 2.05\\
     0.4   & 1.73 & 0.85  & 2.04 \\    
     0.6   & 1.90 & 0.92  & 2.07 \\   
\hline \hline
\end{tabular}
\end{center} \label{tab2}
\end{table}

\subsubsection{The triplet interaction $ \bar{J'} $}

To avoid analyzing the complex three body interactions, the triplet interaction term $ \bar{J'} $ is introduced to count a compacted polarity of triangle unit, which may act similar to the magnetic field $H$. Table \ref{tab3} summarized the $T$s with $ \bar{J'}=-0.3 $, $-0.1$ and $0.1$. It is found that negative $ \bar{J'} $ increases the transition
temperatures or vice versa, and the thermodynamics is very sensitive to $ \bar{J'} $, i.e. a slight variation will dramatically change the overall thermal behaviors. The setup with absolute value of $ \bar{J'} $ larger than 0.3 can converge the system to some bizarre states, which will be detailed in the following section.

\begin{table}[]
\caption{The transition temperature variations with different  $\bar{J'} $.}
\begin{center}
\begin{tabular}{llll} 
\hline\hline $ \bar{J'} $ & $T_{C}$& $T_{K}$ & $T_{C}/T_{K}$\\
\hline -0.3& 1.90 & 0.76 & 2.50 \\
-0.1 & 1.55 & 0.70 & 2.21 \\
0 & 1.33 & 0.69 & 1.93 \\
0.1 & 1.05 & 0.69 & 1.52  \\ 
\hline \hline
\end{tabular}
\end{center} \label{tab3}
\end{table}

\subsection{Two special cases with various $ J' $}

\begin{figure}[htbp]
	\centering{
		\includegraphics[width=0.6\textwidth]{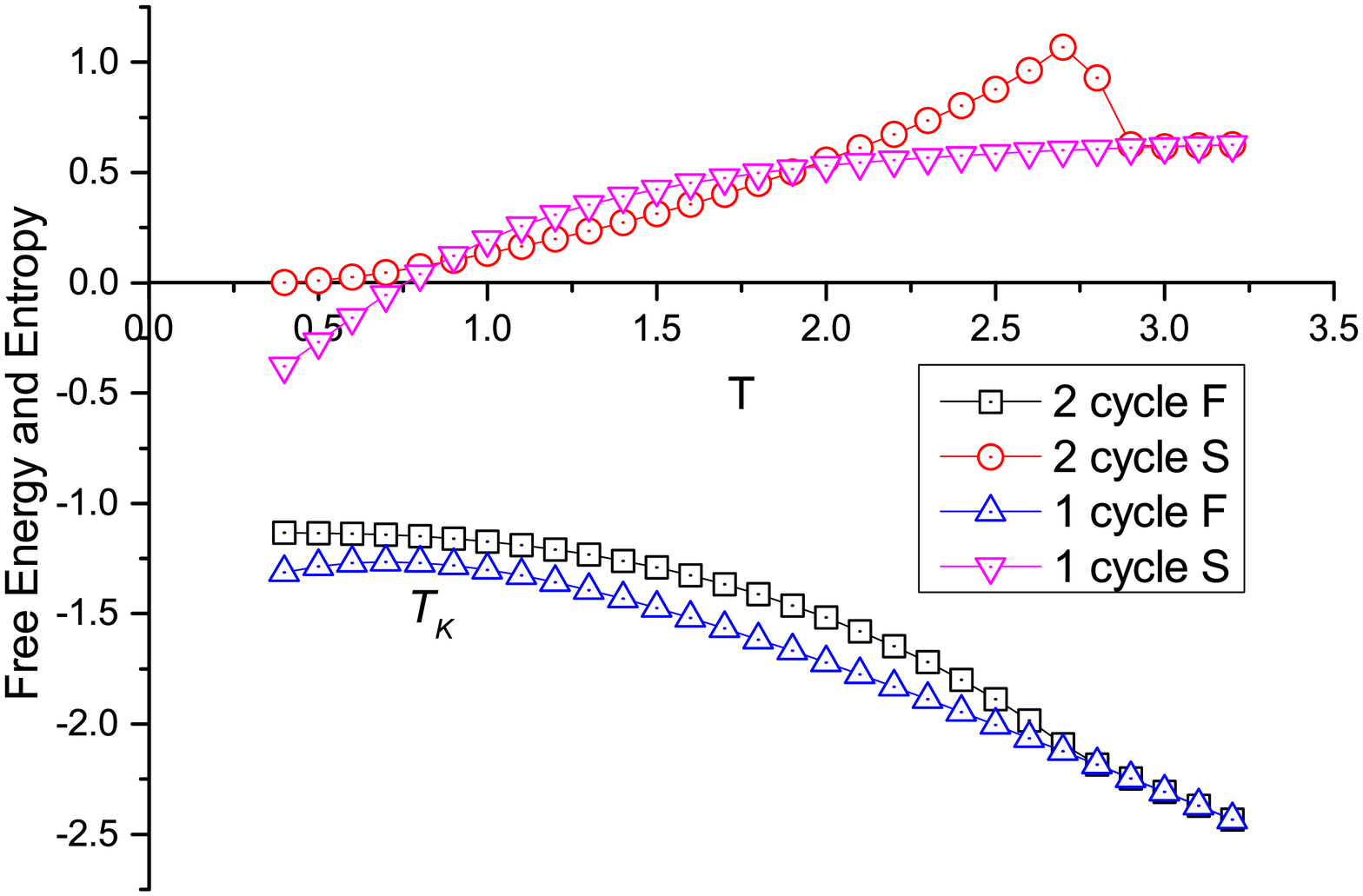}
		\caption{A special case of ZSRL with $\bar{J} = -1 $ and  $ J' = 0.3 $. }
		\label{fig7}
	}
\end{figure}

Since the $ \bar{J'} $ plays a dominant role in ZSRL, the value of $ \bar{J'} $ is limited to be relatively small. Abnormal behaviors can be observed with $ \bar{J'} = 0.3 $ and $ \bar{J'} = -0.5 $. In the first case as shown in Fig. \ref{fig7}, the 2-cycle solution will never have a lower free energy than 1-cycle. Even it has a lower entropy at low temperature, the order-disorder transition cannot be located since there is no cross point. On the other hand, the 1-cycle solution still undergoes Kauzmann paradox. Therefore the only reasonable understanding is that, under this condition, regardless of the thermal state in the bulk, crystal state is not achievable on the surface. With temperature decreasing, we will only have supercooled liquid and the subsequent glassy state. 

\begin{figure}[htbp]
	
	\centering{
		\includegraphics[width=0.6\textwidth]{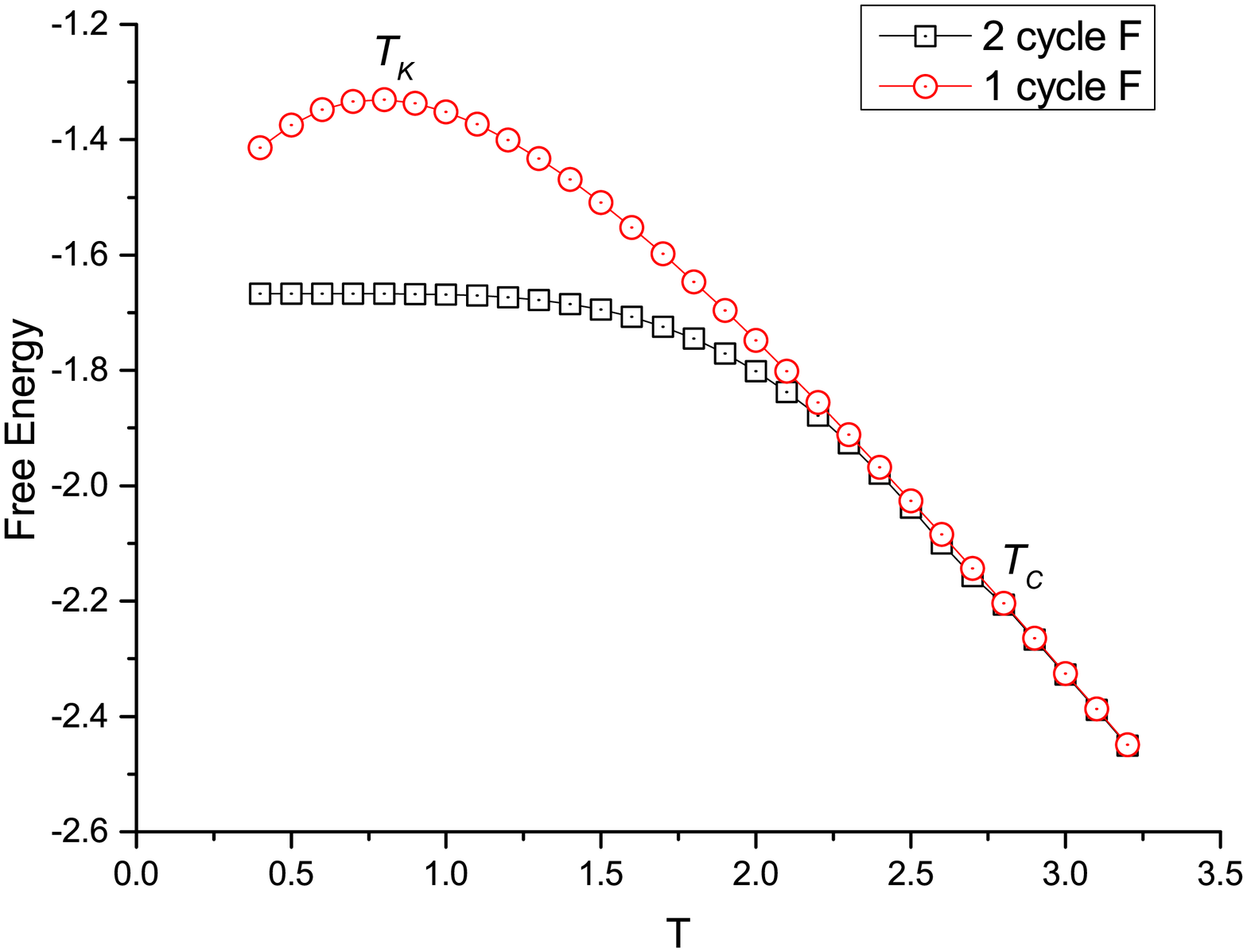}
		
	}

		\centering{
		\includegraphics[width=0.6\textwidth]{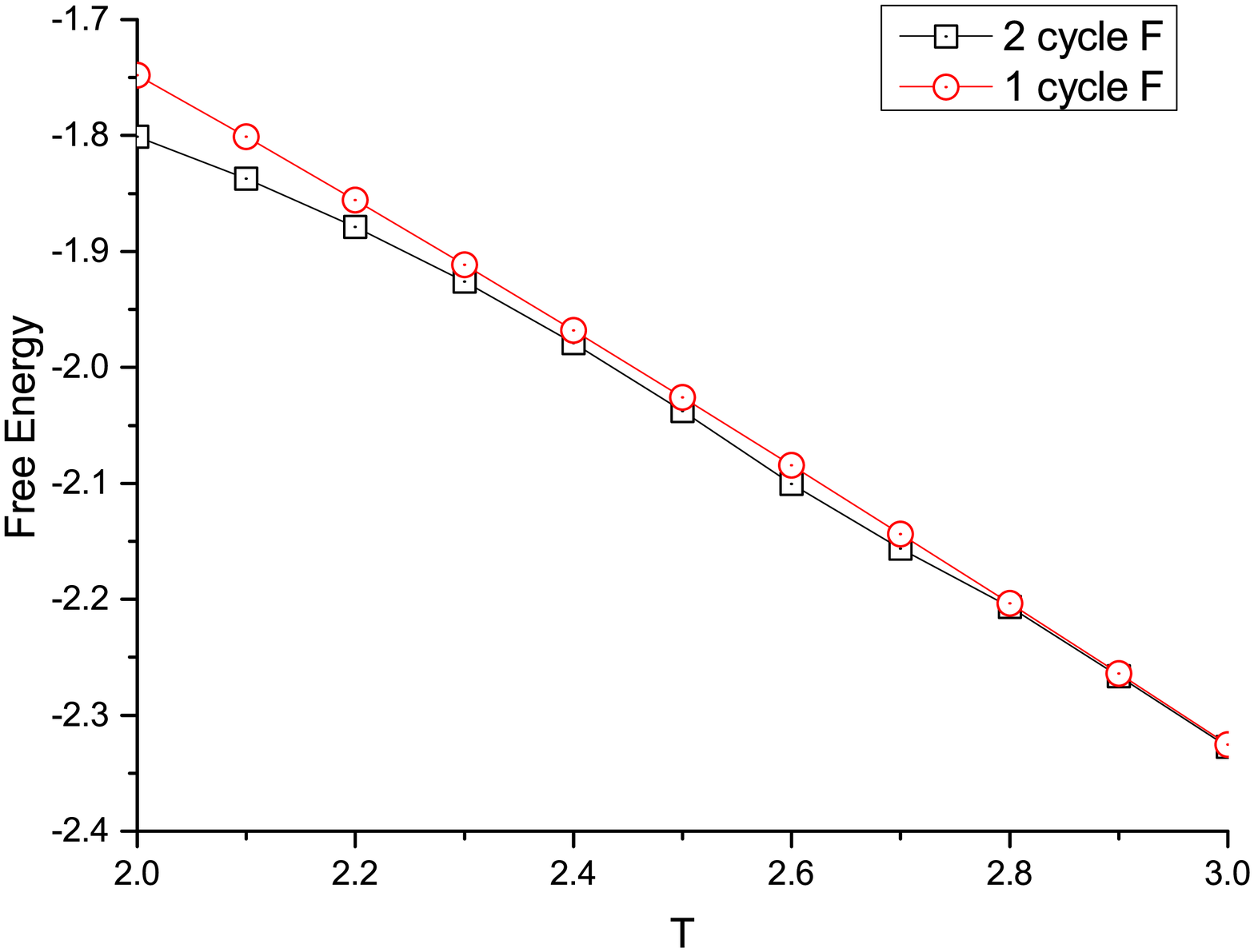}
	
	}
\caption{A special case of ZSRL with $ \bar{J} = -1 $ and  $ J' = -0.5 $: (a) the free energy of 1 and 2 cycle solutions; (b) the enlargement of free energy around $T=2.8$.}\label{fig8}
\end{figure}

The system goes to another extremity for $ \bar{J'} = -0.5 $: Figure \ref{fig8} shows that, below the free energy differentiating point, the 2-cycle solution consistently becomes more stable than 1-cycle, as like the normal behavior of regular antiferromagnetic Ising models, and thereby the order-disorder transition becomes spontaneous magnetization; below $T_C$ the system can either be in crystal ordering or in the metastable supercooled state.  

\section{Conclusion}
A zigzag surface recursive lattice has been constructed to describe a regular square lattice with 1D boundary. The zigzag structure is taken as a surface assembled by triangle units, and halved Husimi trees are hung on the triangle units to represent the bulk portions. With the coordination number of 4 inside the bulk and average 3 on the surface, this model is considered to be a good approximation to a regular square lattice with surface.

The antiferromagnetic Ising model is solved on the lattice, with the constant $x_{B}$ retrieved from regular Husimi lattice to count the bulk contribution, and a uniform solution is obtained on the surface to represent the ordered state. Then the thermodynamics of local area around the origin on surface can be derived by conventional techniques from $\bar{x}$ and $x_{B}$. The transition temperatures are found to be dramatically reduced on the surface comparing which in the bulk, and this reduction is simply due to the dimension downgrade and less interaction constrains on the surface.

The effects of various interaction energy parameters other than nearest-neighbor $\bar{J}$ are investigated. These interactions could either increase or decrease the stability of system and change the transition temperatures according to the Hamiltonian. In addition to the effect of parameters, we have found several interesting behaviors with particular energy setup.

\section{Statements}

\noindent Data Availability: The code and data can be accessed at: https://datadryad.org/review?doi=doi:10.5061/dryad.99t5k4s.

\noindent Competing Interests: We declare no competing interests.

\noindent Authors' Contributions: R.H. designed the model, did the programming, calculation and analysis, and wrote the manuscript. P.D.G. directed the research, derived the theoretical formulation and calculation method, edited the manuscript and approved its submission.

\noindent Funding: This work is financially supported by the National Natural Science Foundation of China (11505110), the China Postdoctoral Science Foundation (2016M591666), and the Taizhou Municipal Science and Technology Program (1701gy15 and 1801gy16).

\noindent Research Ethics: Not applicable.

\noindent Permission to carry out fieldwork: Not applicable.

\noindent Acknowledgements: Not applicable.

\end{document}